\documentclass[preprint2]{aastex}

\begin{document}

 \title{Asymmetric Surface Brightness Distribution of Altair \\
     Observed with the Navy Prototype Optical Interferometer}

\author{Naoko Ohishi\altaffilmark{1} }
 \affil{National Astronomical Observatory of Japan, 
   2-21-1 Osawa, Mitaka, Tokyo 181-8588, Japan }
 \email{naoko.ohishi@nao.ac.jp}

\author{Tyler E. Nordgren}
 \affil{Department of Physics, University of Redlands, 
1200 East Colton Avenue, Redlands, CA 92373
}

\and

 \author{Donald J. Hutter}
 \affil{U. S. Naval Observatory, Flagstaff Station, P. O. Box 1149, 
Flagstaff, AZ 86002-1149}

\begin{abstract}
An asymmetric surface brightness distribution of the rapidly rotating A7IV-V star, Altair, has been measured by the Navy Prototype Optical Interferometer (NPOI). 
The observations were recorded simultaneously using a triangle of three long baselines of 30m, 37m, and 64m, on 19 spectral channels, covering the wavelength range of 520nm to 850nm. 
The outstanding characteristics of these observations are (a) high resolution with the minimum fringe spacing of 1.7mas, easily resolving the 3-milliarcsecond (mas) stellar disk, and (b) the measurement of closure phase which is a sensitive indicator to the asymmetry of the brightness distribution of the source. 
Uniform disk diameters fit to the measured squared visibility amplitudes confirms the Altair's oblate shape due to its rapid rotation. 
The measured observables of Altair showed two features which are inconsistent with both the uniform-disk and limb-darkened disk models, while the measured observable of the comparison star, Vega, are consistent with the limb-darkened disk model. 
The first feature is that measured squared visibility amplitudes at the first minimum do not reach 0.0 but rather remain at $\approx$0.02, indicating the existence of a small bright region on the stellar disk. 
The other is that the measured closure phases show non-zero/180 degrees at all spectral channels, which requires an asymmetric surface brightness distribution. 
We fitted the measured observables to a model with a bright spot on a limb-darkened disk and found the observations are well reproduced by a bright spot, which has relative intensity of 4.7\%, on a 3.38 mas limb-darkened stellar disk. 
Rapid rotation of Altair indicates that this bright region is a pole, which is brighter than other part of the star owing to gravity darkening.
\end{abstract}


\section{Introduction}
%(two features of rapid rotation of star)
The impact of stellar rotation upon fundamental stellar parameters was originally calculated by von Zeipel \citep{vZ1924}. 
He showed that the stellar radius and the surface brightness are expressed as a function of the latitude; the radius increases and the brightness decreases from the pole to the equator. 
%(history of observation)
%(II)
The first attempt to determine the stellar oblateness owing to rapid rotation was performed with the intensity interferometer at Narrabri \citep{RHB1967}. Though they mentioned the possibility of a rotationally flattened shape for Altair, their data was insufficient to decide the oblateness critically. 
%(PTI)
In 2001, van Belle et al (2001) observed Altair using two baselines of the Palomar Testbed Interferometer %\citep[PTI]{MMC1999}
(PTI, Colavita et al. 1999). They calculated the apparent stellar angular diameters from measured squared visibility amplitudes using the uniform-disk model and found that the angular diameters change with position angle. 
%\citep{GTvB2001}. 
This was the first measurement of stellar oblateness owing to rapid rotation. 
They evaluated the effect of surface brightness distribution caused by gravity darkening on their angular diameter data. 
Because their simulation showed that the diameters are changed only a few \%, they neglected the gravity darkening effect. 
%(VLTI)
In 2003, a large oblateness of the Be star Achernar was measured with the VLTI \citep{ADdS2003}. 
They analyzed measured deformation with Roche model which takes the surface brightness distribution owing to gravity darkening into account, but the deformation was too large to be explained within the range of the model. 
%(conclusion from examples)
Though both the measurement at PTI and VLTI demonstrated the determination of the stellar parameters such as orientation of the rotation axis and the apparent oblateness from the squared visibility amplitudes obtained with the interferometer, the effect of surface brightness distribution was not determined. 

Because of the rapid rotation of Altair, where apparent rotational velocities range from $v\sin i$ =190km/s \citep{KGC1984} to 250km/s \citep{TRS1968}, this star is expected to be gravity darkened. 
%(effect on radius)why it was difficult and why we expect we can do it!
However, the effects of the stellar surface brightness distribution on squared visibility amplitudes are small at low-resolution before the first minimum. 
For example, the effect of limb darkening on visibilities becomes more significant after the first minimum. 
Consequently, in order to discuss the surface brightness distribution of rapidly rotating stars, it is indispensable to have high-resolution measurements at least over the first minimum.
%(NPOI)
In 2001, the NPOI achieved an array configuration (see next section) which for the first time allowed for observations with a baseline as long as 64 m. 
With this long baseline and visible observing wavelength, the minimum fringe spacing becomes as small as 1.7 mas at the sky position of Altair. This resolution is high enough to observe this $\approx$3 mas star well over the first minimum. 
%(asymmetry is important)
Moreover, the NPOI is equipped with a system which enables measurement of closure phase in addition to the squared visibility amplitudes since 1996 \citep{JAB1997}.
Closure phase is a sensitive measure to the asymmetry of the brightness distribution of a source. 
If the gravity darkened stellar disk is seen neither equator on nor pole on, the intensity distribution of the source projected on a baseline will be asymmetric and we expect that the measurement of the closure phase gives us useful information about inclination of the star. 
Consequently, we expect that observations of Altair with the NPOI will provide better information than earlier observations in order to examine the surface brightness distribution of this rapidly rotating star. 

%(observation)
Observations are done for four nights from 25 May to 1 June in 2001 using three baselines of 30 m, 37 m, and 64 m of NPOI. 
We examine the apparent stellar diameters reduced using a uniform-disk model, and discuss the surface brightness distribution of this star. 

\section{Observations}
\begin{deluxetable}{llllll}
\tablewidth{0pt}
\tablecaption{\label{TAB1}Arrangement of three baselines used for observation.}
\tablehead{
\colhead{baseline}&\colhead{station}&\colhead{E [m]}&\colhead{N [m]}
&\colhead{Z [m]}&\colhead{length [m]}
}
\startdata
OB1&AW-AE&-37.379245&-2.60124&0.000314&37.46964647\\
OB2&W7-AW& -23.813216&-17.49076&-0.080794&29.54661519\\
OB3&AE-W7& 61.192461&20.092&0.08048&64.40661631\\
\enddata
%\tablecomments{a}
\end{deluxetable}
%NPOI
The NPOI, located near Flagstaff, AZ, is a long-baseline optical interferometer which consists of an astrometric subarray (four elements) and an imaging subarray (six elements) \citep{AJT1998}. 
For each baseline, data are collected in thirty-two spectral channels cover the observing wavelength from 450nm to 850nm. 
%
%baseline
We used three baselines consisting of three siderostats; two of which were in astrometric subarray, AW, AE, and one of which was in imaging subarray, W7. 
In this paper, we state each baseline as follows: OB1(AW-AE), OB2(W7-AW), and OB3(AE-W7). 
The arrangement and the length of each baseline is shown in Tab. \ref{TAB1}. 

%time
We observed Altair for four nights from 25-27 May, and 1 June in 2001 and obtained
19 scans \citep{CAH1998} of data. 
Detailed observation table is shown in Tab. \ref{TAB2}.
\begin{table*}
\begin{center}
\caption{\label{TAB2} Observation table for Altair. }
%\begin{deluxetable}{llccccccc}
%\tablewidth{0pt}
%\tablecaption{\label{TAB2} Observation table. Position angle 
%is written as pa; 180 degree was added for OB3.}
%\tablehead{
%\colhead{}&\colhead{}&\colhead{}&\colhead{OB1}&\colhead{}&
%\colhead{OB2}&&OB3&\\
\begin{tabular}{llccccccc}
\tableline\tableline
&&&OB1&&OB2&&OB3&\\
date&scan&Hour Angle& P. A.& UDD&P. A.& UDD &P. A.& UDD\\
    &  no.  & [hr]  & [deg] & [mas] & [deg] & [mas] & [deg] & [mas]\\
%\startdata
\tableline
May25& Pt 81 &-0.470 & 182.5 & 3.10 & 211.6 & 3.23 & 195.2 & 3.16 \\
     & Pt 84 &-0.318 & 182.8 & 3.01 & 212.1 & 3.16 & 195.6 & 3.18 \\
     & Pt 87 &-0.167 & 183.2 & 2.99 & 212.7 & 3.12 & 196.0 & 3.20 \\
     & Pt 90 &-0.017 & 183.5 & 2.97 & 213.4 & 3.09 & 196.4 & 3.19 \\
     & Pt 93 & 0.132 & 183.9 & 3.08 & 214.1 & 3.17 & 196.8 & 3.15 \\
     & Pt 97 & 0.629 & 185.1 & 3.14 & 216.7 & 3.44 & 198.5 & 3.18 \\
     & Pt 100& 0.827 & 185.6 & 3.16 & 218.0 & 3.32 & 199.3 & 3.18 \\
\hline
May26& Pt 30 &-0.242 & 183.0 & 3.04 & 212.4 & 3.20 & 195.8 & 3.17 \\
     & Pt 35 & 0.208 & 184.1 & 2.91 & 214.4 & 3.13 & 197.1 & 3.18 \\
     & Pt 38 & 0.775 & 185.5 & 3.20 & 217.6 & 3.44 & 199.1 & 3.19 \\
\hline
May27& Pt 23 &-0.967 & 181.4 & 3.15 & 210.0 & 3.25 & 194.1 & 3.17 \\
     & Pt 30 &-0.484 & 182.5 & 3.14 & 211.5 & 3.25 & 195.2 & 3.14 \\
     & Pt 35 &-0.146 & 183.2 & 3.03 & 212.8 & 3.18 & 196.0 & 3.21 \\
     & Pt 39 & 0.142 & 183.9 & 3.08 & 214.1 & 3.28 & 196.9 & 3.18 \\
     & Pt 44 & 0.475 & 184.7 & 3.27 & 215.8 & 3.54 & 198.0 & 3.17 \\
\hline 
Jun01& Pt 14 &-1.041 & 181.2 & 3.18 & 209.8 & 3.26 & 194.0 & 3.12 \\
     & Pt 19 &-0.726 & 181.9 & 3.09 & 210.7 & 3.31 & 194.6 & 3.15 \\
     & Pt 24 &-0.410 & 182.6 & 3.06 & 211.8 & 3.23 & 195.3 & 3.17 \\
     & Pt 29 &-0.120 & 183.3 & 3.03 & 212.9 & 3.23 & 196.1 & 3.16 \\ 
%\enddata
\tableline
\end{tabular}
\tablenotetext{a}{
Position Angle, P. A., is east of north. For OB3, 180 degree are added. UDD is uniform disk diameter. }
\end{center}
\end{table*} 
%(comparison)
Interposed with the observations of Altair were observations of the calibration star, $\zeta$ Aql, and a comparison star $\alpha$ Lyr, Vega. 
We chose Vega as a comparison star because of its similar spectral type, magnitude, and angular size to that of Altair. 
In addition, Vega is a good comparison when we discuss the oblateness of Altair because we expect negligible deformation from a circularly symmetric disk given that its observed rotational velocity is less than 20km/s \citep{RFF1983}. 
During our observations, the measured squared visibility amplitudes of Vega showed dependence on zenith angle when the star passed within 10 degrees of zenith. No such dependence was found outside this region.
Consequently, we removed data of Vega with zenith angle smaller than 10 degree. Fourteen scans of Vega remained after this selection. 
We used all 19 scans of Altair because that star doesn't pass the region close to the zenith.
We also used all data of $\zeta$ Aql and the number of scans of this star during four nights was 24. 
%

%revision1_1 start
$\zeta$ Aql was used to calibrate measured observables of both Altair and Vega. 
The uniform disk diameter of this calibration star is computed as 0.80mas from its observed color and magnitude \citep{DM1991}. 
This value is also used in other observation \citep{TEN1999}. 
Though the size of the star, 0.80mas, is small compared with Vega and Altair, the star is partly resolved with the longest baseline; measured with the full 64.4m baseline, squared visibility amplitudes become about 0.7 for a perfect interferometer. 
Because $\zeta$ Aql is a rapidly rotating star with $v \sin i\sim 320$km/s \citep{FR2002}, we examined the possibility that using this star as a circular disk with the diameter of 0.80mas biases the results of Vega and Altair. 
We evaluated the effect of rotation as 5\% changes in diameter and 5\% asymmetric brightness distribution on the star owing to gravity darkening. 
At the longest baseline, we found that these features change the squared visibility amplitudes about 6\% and phase about 3 degree at most. 
The calculated possible errors of squared visibility amplitudes and phase were almost the same level with the measurement errors, which are described in sec. 2.2. 
Consequently, the effect of the rapid rotation of the calibration star is not critical to our result, but we need to be careful when we reduce precise physical parameters directly from measured observables. 
%revision1_1 ... end

%channel
Because the signal to noise ratio of the measured visibilities decreases as the wavelength becomes shorter, we used only the 20 reddest of the 32 spectral channels, thus covering wavelengths from 520nm to 850nm. 
%revision1_2 ... start
The raw squared visibility amplitudes of the calibration star depend on channels and baselines. The highest raw $V^2$ was $0.71\pm0.04$ at the reddest channel of OB2 ($V^2$ for a perfect interferometer on a 0.8mas star is 0.96) and the lowest raw $V^2$ was 0.13$\pm$0.01 at the bluest channel of OB3 ($V^2$ for a perfect interferometer on a 0.8mas star is 0.6). 
%revision1_2 ... end
We also removed the data obtained at the spectral channel with wavelength, $\lambda$ = 633nm, because this channel contains light from the NPOI's He-Ne metrology laser. 
One detector for OB3 was not available during our observations and one other one didn't work on 1st June 2001. 
Consequently, the number of data were, 19 spectral channels $\times$ 19 scans=361 squared visibility amplitudes on OB1 and OB2, 18 spectral channels $\times$ 15 scans + 17 spectral channels $\times$ 4 scans = 338 squared visibility amplitudes on OB3, triple amplitudes, and closure phases. 
The total number of data were 361 $\times$ 2 + 338 $\times$ 3 = 1,738. 
\subsection{Detector dead time correction}
Before calibrating the measured observables, we corrected the effect of detector dead time. 
When we observe bright stars such as Altair and Vega, the effect of detector deadtime $t_{\rm d}$ is not negligible.
The deadtime of the detectors, which are used at the NPOI, are approximately 200ns.
The number of photons counted per unit cycle (500Hz) at the detectors of the red spectral channels become about 800; one photon is detected per approximately 2.5 $\mu$ sec.
In such case, the number of photons counted at the detectors becomes less than the number of incident photons, and the response of the detector becomes non-linear. 
This is a well known effect and we calculate the actual photon count rate, $CR_{\rm a}$, using measured photon count rate, $CR_{\rm m}$, as follows,
\begin{equation}
\label{eq:s2_CRa}
CR_{\rm a}=\frac{CR_{\rm m}}{1-t_{\rm d}CR_{\rm m}}.
\end{equation}
Here, count rate is the number of photons counted per one second.

We calculated the actual visibility $|V_{\rm a}|$ using the measured visibility $|V_{\rm m}|$ as follows,
\begin{equation}
\label{eq:s2_Va}
|V_{\rm a}|=\frac{|V_{\rm m}|}{1-t_{\rm d}CR_{\rm m}
\times(1-|V_{\rm m}|^2)}.
\end{equation}
This correction changed the squared visibility amplitudes of Vega up to 20\%. 
In this paper, we treat the detector deadtime correction as described above. 
Triple amplitudes are also corrected based on eq. (\ref{eq:s2_Va}). 

\subsection{Measurement errors}
%error
The statistical errors of measured quantities are calculated and recorded for each scan. 
However, actual measurement errors are sometimes dominated by long-term errors rather than statistical errors within short-term measurement \citep{MW2001}. 
Generally, it is not easy to evaluate long-term systematic errors.  
In this paper, we calculated the variance of calibrated quantities, $\sigma_{\rm cal}(\lambda_{j})$, of the calibration star as long-term errors and estimated errors of measured squared visibility amplitudes, $\sigma_{\rm vs}(t_{i}, \lambda_{j})$, as follows,
\begin{eqnarray}
\label{eq:s2_sVS}
\lefteqn{\sigma^2_{\rm vs}(t_{i}, \lambda_{j})
=
\sigma^2_{\rm vs, stat}(t_{i}, \lambda_{j})\nonumber}\\
&&+\sigma^2_{\rm vs, cal}(\lambda_{j})
\times \frac{|V(t_{i}, \lambda_{j})|^2}
{\sum_{i=0}^{N-1}|V_{\rm cal}(t_{i}, \lambda_{j})|^2/N}. 
\end{eqnarray}
Here, $t_{i}$ is the time at the i-th scan, $\lambda_{j}$ is the wavelength at the j-th spectral channel, $\sigma_{\rm vs, stat}(t_{i}, \lambda_{j})$ is the statistical error, $|V(t_{i}, \lambda_{j})|^2$ is the calibrated squared visibility amplitudes of the source, $|V_{\rm cal}(t_{i}, \lambda_{j})|^2$ is the calibrated squared visibility amplitudes of the calibration star, and $N$ is the number of scans.
We also estimated measurement errors of triple amplitudes as above.
For closure phases, we used the following equation, 
\begin{equation}
\label{eq:s2_sCP}
\sigma^2_{\rm cp}(t_{i}, \lambda{j})
=%\sqrt{
\sigma^2_{\rm cp, stat}(t_{i}, \lambda_{j})
+\sigma^2_{\rm cp, cal}({\rm \lambda}).
%}
\end{equation}
The ratio of errors calculated using eqs. (\ref{eq:s2_sVS}), and (\ref{eq:s2_sCP}) to the statistical errors were 4.8 (for squared visibility amplitudes at OB1), 15.4 (OB2), 1.4 (OB3), 2.7 (for triple amplitudes), 1.1 (closure phases). 
This means that the measurement errors of these observations are nearly dominated by the long-term errors rather than short-term statistical errors. 
Even if the measurement errors are dominated by long-term errors, it is confirmed that data obtained at NPOI shows no systematic error compared with data obtained at other interferometer \citep{TEN2001}. 
\section{Results}
The squared visibility amplitudes on three baselines, triple amplitudes, and closure phases on about 19 channels were obtained simultaneously for each scan in our observation. 
Visibility measured at a baseline with projected baseline, $B_{\rm p}$, of a stellar interferometer, which is the Fourier transform of the brightness distribution of the source, is written as follows, 
\begin{equation}
\label{eq:s3_V}
V(kB_{\rm p})=\frac{\int I(x)\exp[-ikB_{\rm p} x] {\rm d} x}
{\int I(x) {\rm d} x},
\end{equation}
where $I(x)$ is the brightness distribution projected to the baseline, $k=2\pi/\lambda$ is the wavenumber. 
%, $B_{\rm p}$ is the projected baseline.
%
For example, visibility of the uniform-disk with the angular radius of $r_{\rm UD}$ is written as follows, 
\begin{equation}
\label{eq:s3_V2UD}
V_{\rm UD}(r_{\rm UD})%(kB_{\rm p};r_{\rm UD})
=\frac{2J_1(k B_{\rm p} %\theta_{\rm UD}/2
r_{\rm UD})}
{k B_{\rm p} %\theta_{\rm UD}/2
r_{\rm UD}}. 
\end{equation}
Visibility of the limb-darkened disk with the angular radius of $r_{\rm LD}$ and the linear limb darkening coefficient of $u(k)$ is written as follows \citep{AQ1996},  
\begin{eqnarray}
\label{eq:s3_V2LD}
\lefteqn{V_{\rm LD}(r_{\rm LD}) 
=\frac{6}{3-u(k)}%\left\{
\{1-u(k)\}\frac{2J_1(k B_{\rm p} %\theta_{\rm LD}/2
r_{\rm LD})}{k B_{\rm p} %\theta_{\rm LD}/2
r_{\rm LD}} } \nonumber\\
&&+\frac{6}{3-u(k)} u(k) \sqrt{\frac{\pi}{2}}
\frac{J_{3/2}(k B_{\rm p} %\theta_{\rm LD}/2
r_{\rm LD})}{(k B_{\rm p} %\theta_{\rm LD}/2
r_{\rm LD})^{3/2}}
%\right\}
. 
\end{eqnarray}

The triple product is the product of the visibilities on three baselines that form a triangle, 
\begin{equation}
\label{eq:s4_TP}
V_{\rm TP}=|V_1|\exp(-i\Phi_1) |V_2| \exp(-i\Phi_2) |V_3| \exp(-i\Phi_3). 
\end{equation}
The triple amplitude is the absolute value of the triple product, 
\begin{equation}
\label{eq:s4_TA}
|V_{\rm TP}|=|V_1|%\times
|V_2|%\times
|V_3|, 
\end{equation}
and the closure phase is the phase of the triple product,  
\begin{equation}
\label{eq:s4_CP}
\Phi_{\rm c}=\Phi_1+\Phi_2+\Phi_3.
\end{equation}
As we can see from eq. (\ref{eq:s3_V}), if the brightness distribution of the source projected to a baseline, $I(x)$, is symmetric with $x$, the imaginary part of the visibility is zero and the phase of the visibility takes 0 or $\pm 180$ degree. 
If the brightness distribution of the source is asymmetric, the imaginary part of the visibility becomes non-zero and the phase takes non-zero/$\pm 180$ degree.
Generally, it is not easy to measure phase of visibility with a ground interferometer because of atmospheric turbulence.
However, the closure phase cancels the effect of atmospheric turbulence and conserves information of the source. 
Consequently, closure phase is a useful interferometric observable in order to discuss asymmetry of brightness distribution of the source. 

\subsection{Measured squared visibility amplitudes, triple amplitudes, and closure phases}
\begin{figure*}
\epsscale{1.80}
\plotone{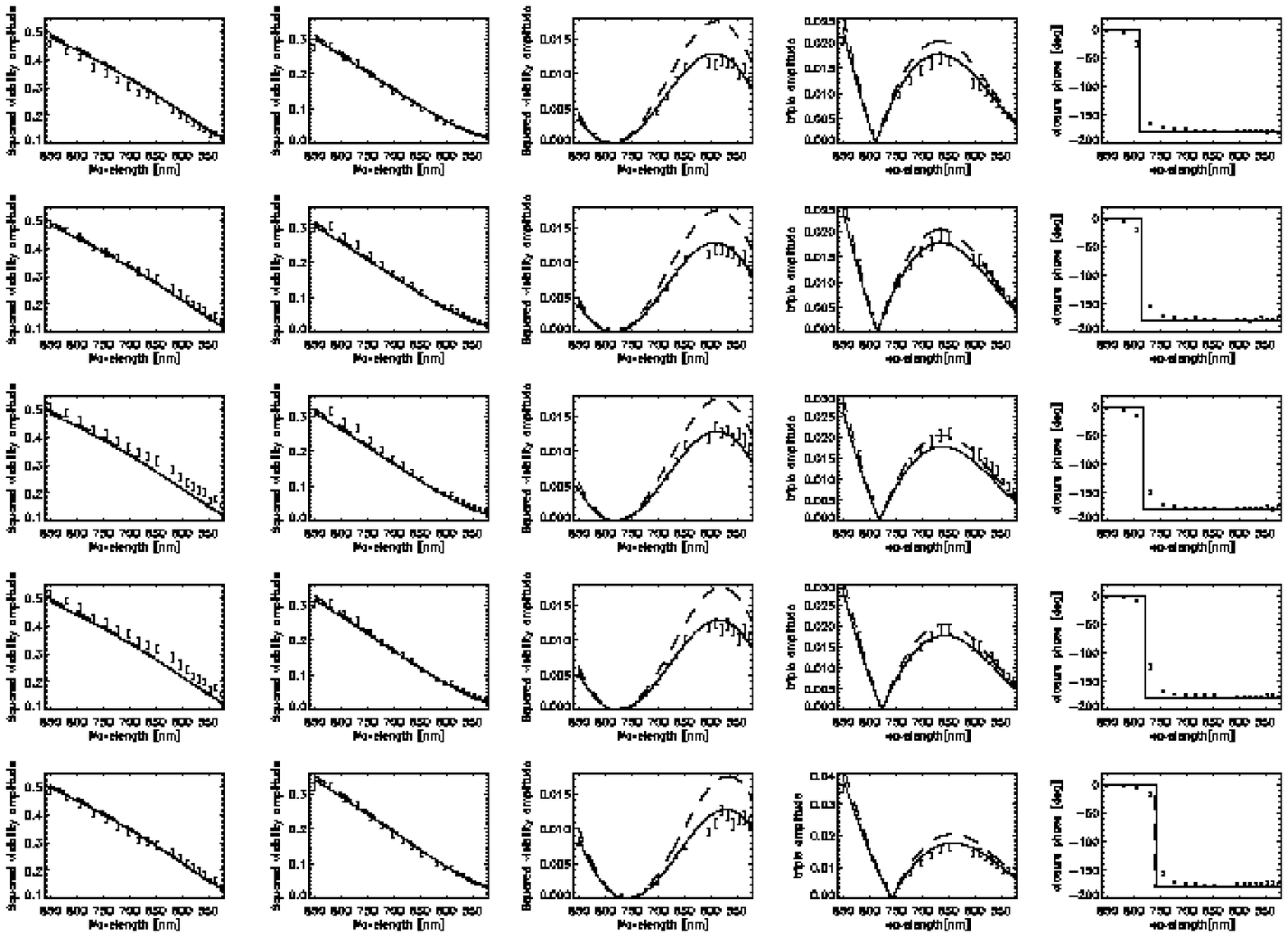}
\plotone{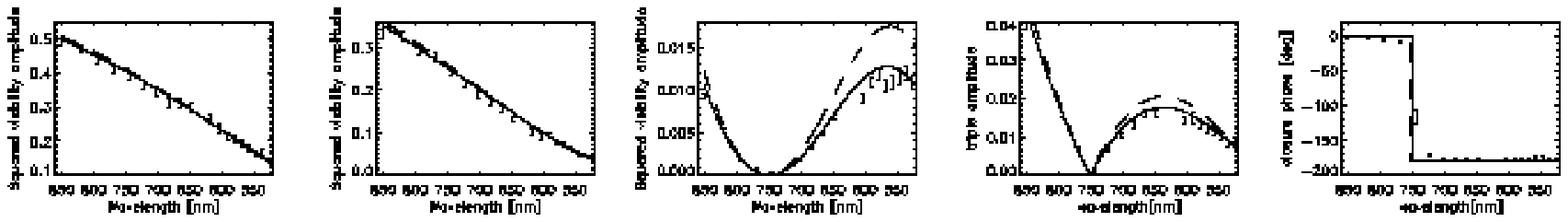}	%!!T
\vspace{-7.5cm}		%!!T
\caption{\label{fig1}
From left to right, squared visibility amplitudes on OB2, OB1, and OB3, triple amplitudes, and closure phases of Vega measured during 25 May 2001. Six scans were obtained. 
Dashed lines show the uniform-disk model with diameter of 3.11mas. 
Solid lines show the limb-darkening model with diameter of 3.22mas. 
We used linear limb darkening coefficients calculated by van Hamme, with $T_{\rm eff}$=9500K and log g=4.0.
The limb-darkening model reproduces the measured squared visibility amplitudes and triple amplitudes better than the uniform disk model. }
\end{figure*}

\begin{figure*}
%\epsscale{.80}
\plotone{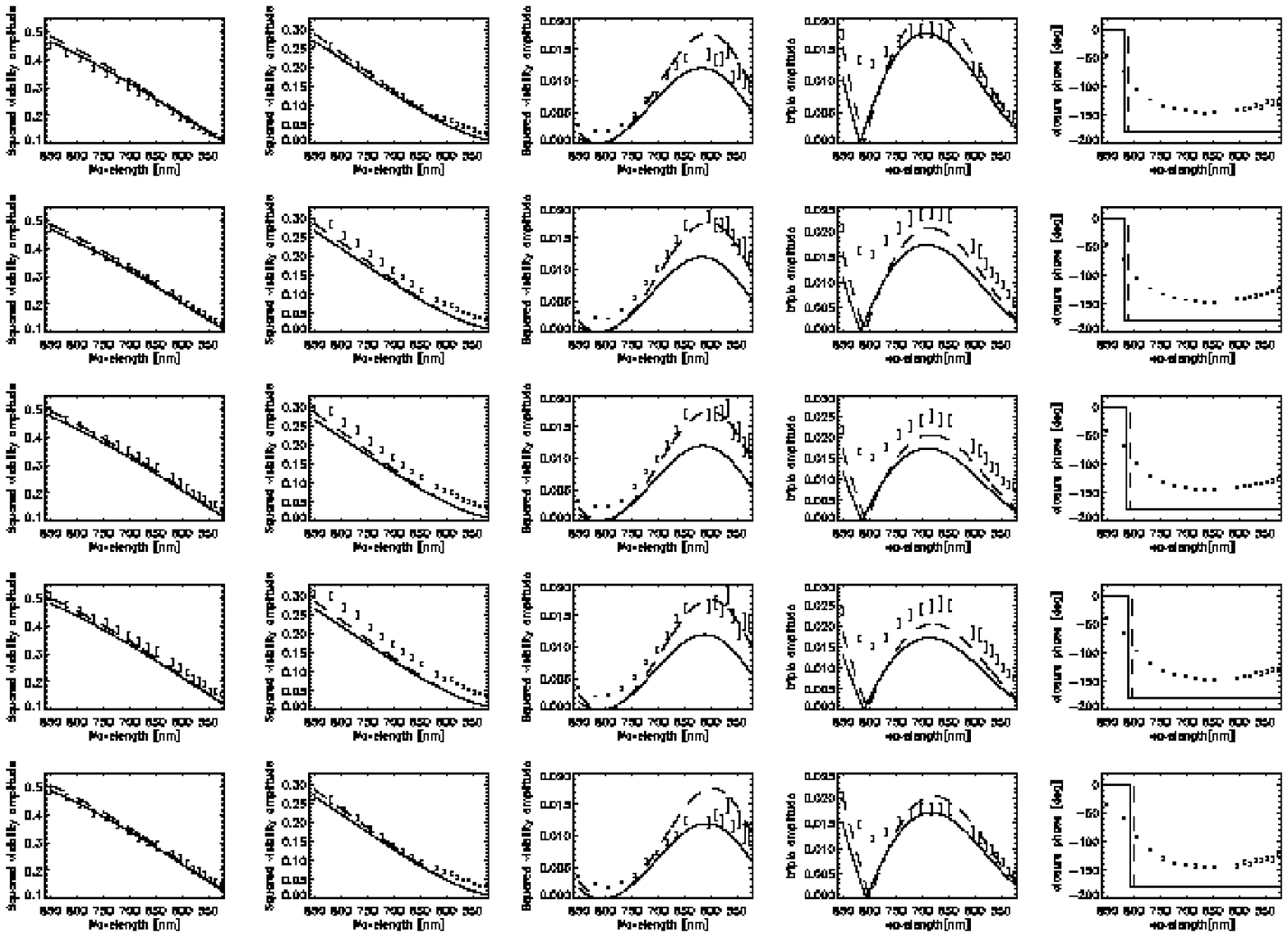}
\plotone{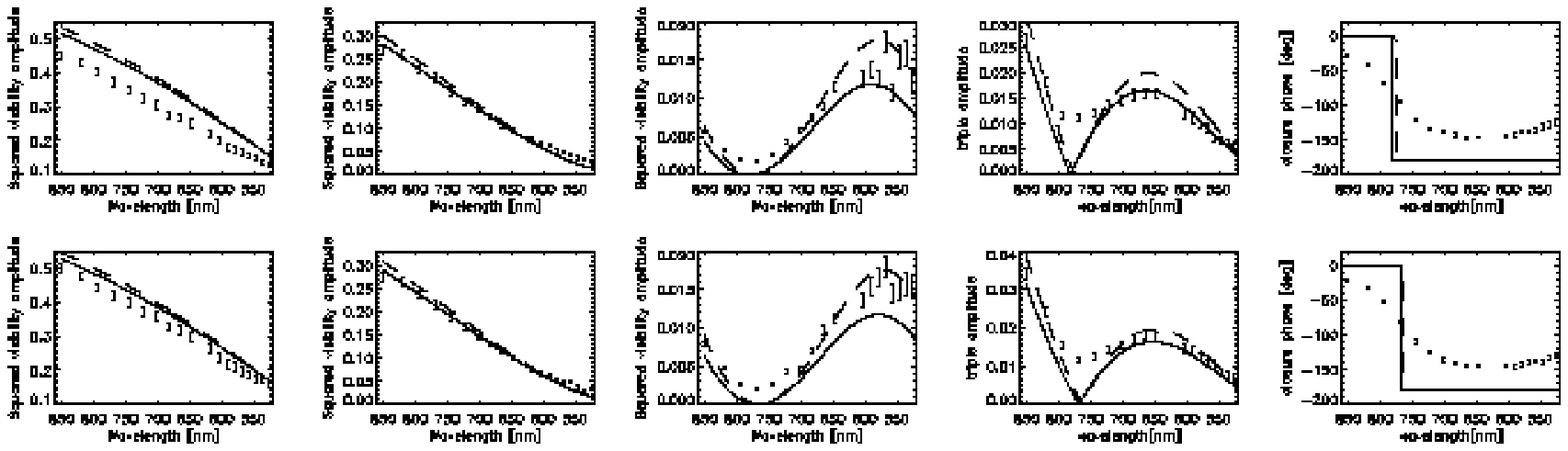}	%!!T
\vspace{-5cm}		%!!T
\caption{\label{fig2}
From left to right, squared visibility amplitudes at OB2, OB1, and OB3, triple amplitudes and closure phases of Altair measured during the night of 25 May 2001. Seven scans were obtained.
Dashed lines show the uniform-disk model with diameter of 3.17mas. 
Solid lines show the limb-darkening model with diameter of 3.32mas. 
We used linear limb darkening coefficients calculated by van Hamme, with $T_{\rm eff}$=7750K and log g=4.0. 
It is significant that the squared visibility amplitudes at OB3 and the triple amplitudes around the first minimum don't decrease to zero, and the closure phases show non-zero/180 degree at all spectral channels. 
}
\end{figure*}

%Vega
Vega was chosen as a comparison star due to its similarities in size, spectral type, and magnitude with Altair, but yet without the high observed rotational velocity which would lead to apparent oblateness. 
Figure \ref{fig1} shows the squared visibility amplitudes and triple products measured during the night of 25 May 2001 for Vega. 
%Altair
Figure \ref{fig2} shows the observables of Altair measured on the same night. 
The dashed lines show the uniform-disk model with the fitted angular diameters of $2r_{\rm UD}=3.17$mas for Altair, and $2r_{\rm UD}=3.11$mas for Vega. 
The solid lines show the limb-darkened disk model with the fitted angular diameters of $2r_{\rm LD}=3.32$mas for Altair and $2r_{\rm LD}=3.22$mas for Vega. 
In order to calculate the limb darkening model, we used linear limb darkening coefficients calculated by van Hamme \citep{WVH1993} with parameters, $T_{\rm eff}$=7750K and log g=4.0 for Altair, and $T_{\rm eff}$=9500K and log g=4.0 for Vega. 

The measured observables of Vega are better fitted with the limb-darkened model (reduced $\chi^2=7.5$), than with the uniform-disk model (reduced $\chi^2=$17.2). 
The limb-darkened disk diameter of Vega measured at PTI was $3.28\pm0.01$mas \citep{DRC2001} and the diameters are consistent with about 2\% error. 
However, Altair was not well fitted with either the uniform-disk model (reduced $\chi^2=150$) or the limb-darkened model (reduced $\chi^2=154$). 
Altair's large inconsistency between the measured observables and models is mainly owing to the facts that: (a) the squared visibility amplitudes at OB3 and the triple amplitudes around the first minimum do not decrease to zero, and (b) the closure phases show non-zero/180 degree at all spectral channels. 
In addition, the measured squared visibility amplitudes at OB1 are slightly bigger than the circular uniform disk model while that at OB2 is slightly smaller than the model. 
The discrepancy of measured squared visibility amplitudes and triple amplitudes from the model around the first minimum indicates that there is a small bright component on the stellar disk, which is not resolved even with the longest baseline. 
Non-zero/180 degree of closure phases means the brightness distribution of the source is asymmetric. 

\subsection{Apparent Stellar Diameters Reduced from the Uniform Disk Model}
%model
Before considering the discrepancy from the uniform-disk model, we first examine whether the uniform disk angular diameter reduced from the squared visibility amplitudes changes with position angle and whether the change is consistent with the PTI result. 
We consider the diameter change assuming the elliptical shape of the stellar disk. 
The brightness distribution of an uniform ellipse projected to a baseline with position angle $\phi$ becomes the same with that of the circular uniform-disk with angular radius, $r(\phi)$, 
\begin{equation}
\label{eq:s3_rphi}
%\theta_{\rm UD}
r(\phi)=\sqrt{a^2\sin^2(\phi-\phi_0)+b^2\cos^2(\phi-\phi_0)}, 
\end{equation}
where $a$ is the angular radius at the major axis, $b$ is the angular radius at the minor axis, and $\phi_0$ is the orientation angle of the ellipse on the sky, where $\phi_0=0$ corresponds to the minor axis pointing to the north on the sky. 
As a result, the squared visibility amplitude becomes the same as that of eq. (\ref{eq:s3_V2UD}) with the angular radius, $r(\phi)$. 
Consequently, in order to determine the parameters of the ellipse, we reduce the data using the following procedure: first, calculate the angular diameters at a position angle by fitting the squared visibility of each scan at each baseline with eq. (\ref{eq:s3_V2UD}), then compare the reduced diameters with eq. (\ref{eq:s3_rphi}). 
Notice $r(\phi)$ is a little different from the intercept of an ellipse at an angle $\phi$. 

%error
Though the errors of fitted angular diameters are calculated from the errors of visibilities, we found that the scatters of the diameters were a few times larger than the calculated errors at OB1 and OB2. 
Thus we define the errors of the fitted angular diameters in another way. 
Considering the small change of position angle at each baseline, $4^{\circ}$ (OB1), $5^{\circ}$ (OB2), $8^{\circ}$ (OB3) (Tab. \ref{TAB2}) during the observation, we regard the change in angular diameter of the ellipse with position angle as small compared to the scatter of the fitted angular diameters.
We therefore used the variance of the fitted angular diameters at each baseline, 0.09mas (OB1), 0.11mas (OB2), 0.02mas (OB3) as errors of the fitted angular diameters. 
The error of the fitted angular diameters on OB3 is smaller compared with that of on OB1 and on OB2. 
We consider this fact is owing to that the measured visibilities cover the first minimum at OB3 and the position of the first minimum improves the precision of the fitted diameters. 

\begin{figure*}
\epsscale{2.1}
\plotone{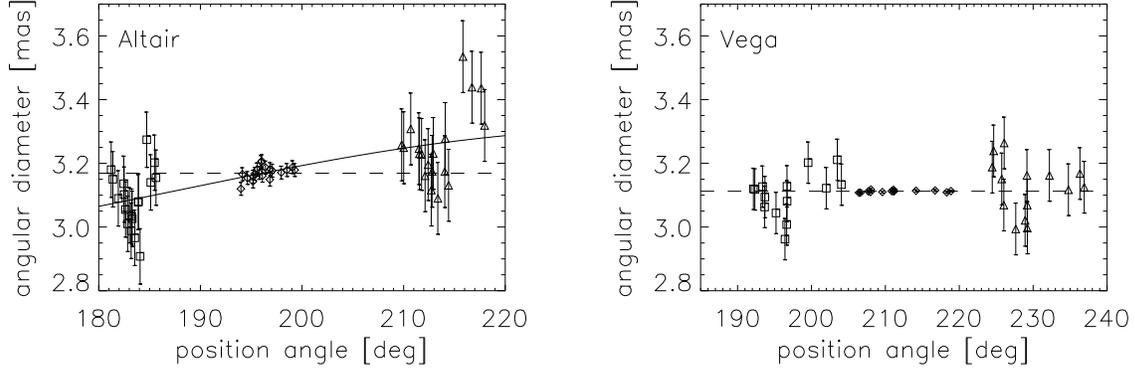}
\vspace{-5.5cm}
\caption{\label{fig3} 
Uniform-disk diameters of Altair (left) and Vega (right) as a function of baseline position angle. 
Squares show fitted angular diameters at OB1, diamonds show fitted angular diameters at OB3, and triangles show fitted angular diameters at OB2.  
%Calculated error bars %from eq. \ref{eq_er} are 
%are smaller than the size of mark.
Dashed lines show fitted angular diameters of 
%$2r_{\rm UD}=
3.17mas (Altair) and 
%$2r_{\rm UD}=
3.11mas (Vega). 
Solid line shows fitted ellipse with parameters of $2a$=3.31mas, $2b$=2.93mas, and $\phi_0=35^{\circ}\pm18^{\circ}$. 
}
\end{figure*}

Figure \ref{fig3} shows fitted angular diameters with horizontal axis of position angle. 
The reduced $\chi'^2$ of 1.5 calculated based on the scatter of the diameters of Altair, with the circular uniform disk model is improved to 0.95 
%with the diameter, $2r_{\rm UD}$=3.17 $\pm$ 0.02mas 
when we allow the diameter of Altair to change with position angle. 
Fitted parameters with eq. (\ref{eq:s3_rphi}) become $2a=3.31\pm0.09$mas, $2b=2.93\pm0.17$mas, $\phi_0=35^{\circ}\pm18^{\circ}$. 
In our case, the uniform disk model estimates the angular diameters of Altair about 0.14mas smaller than the limb-darkening model. 
Resultant parameters corrected for the limb darkening effect are consistent with the PTI result \citep{GTvB2001}, $2a=3.461\pm0.038$mas, $2b=3.037\pm0.069$mas, $\alpha_0=25^{\circ}\pm9^\circ$. 
%Vega
Compared with Altair, fitted diameters and errors calculated from the scatters of the fitted diameters on each baseline of Vega are shown in Fig. \ref{fig3}. 
The diameter fitted with the circular uniform-disk model was, $2r_{\rm UD}=3.11\pm0.01$mas, $\chi'^2=1.0$ and the value of $\chi'^2$ was not changed when we used the elliptical model. 
%
%Parameters of the fitted ellipse were, $2a=3.11$mas, $2b=2.90$mas, $\phi_0=56^{\circ}$, but $\chi'^2=1.0$ was not improved. 
%
We found that Altair is well explained by the ellipse rather than the circular disk while Vega is well explained as the circular disk. 
The dependence of fitted diameters of Altair on the position angle was similar to PTI's. 

\subsection{A small bright region on the limb darkening disk}

In this section, we examine whether the two features of measured observables of Altair which are inconsistent with both the uniform-disk and the limb-darkened disk models are reproduced with the model of a single bright region on the limb-darkened disk. 
When a bright spot with relative intensity of $I_{\rm p}$ was located at ($r_{\rm p}$, $\phi_{\rm p}$) in polar coordinate on a limb-darkening disk with angular radius of $r_{\rm s}$, visibility is written as follows, 
\begin{eqnarray}
\label{eq:s5_model1}
%V_{\rm model1}(k, B_{\rm p}, I_{\rm p}, r_{\rm s}, r_{\rm p}, \phi_{\rm p})
\lefteqn{V_{\rm model}(I_{\rm p}, r_{\rm s}, r_{\rm p}, \phi_{\rm p})
=(1-I_{\rm p})V_{\rm LD}(r_{\rm s})}\nonumber\\
&&+I_{\rm p}\exp\left\{-i k B_{\rm p} 
r_{\rm p}\cos(\phi%_{\rm pa}
-\phi_{\rm p})
\right\},
\end{eqnarray}
With this model, the visibility amplitude at the first minimum becomes $I_{\rm p}$. 
We can see the value of squared visibility amplitudes around the first minimum about 0.02 and expect that $I_{\rm p}=0.04\sim 0.05$. 
Though the existence of bright spot changes stellar diameter $2r_{\rm s}$, we expect that the change of $2r_{\rm s}$ from $2r_{\rm LD}$=3.32mas is small. 
We searched the optimal position of the spot ($r_{\rm p}$; 0$\sim r_{\rm s}$, $\Delta r_{\rm p}$=0.01$r_{\rm s}$, $\phi_{\rm p}$; 0$\sim$360, $\Delta \phi$=1 [deg]) for each set of ($I_{\rm p}$, 2$r_{\rm s}$) ranging ($I_{\rm p}$; 0.034$\sim$0.06, 2$r_{\rm s}$; 3.25$\sim$3.51mas). 
As a result, we found that the set of parameters, ($I_{\rm p}$=0.047, $r_{\rm s}$=3.38mas) gives the best $\chi^2_{\rm c}$ of 7.3. 
%based on errors eqs. (\ref{eq:s2_sVS}), (\ref{eq:s2_sCP}). 
%
The $\chi^2$ value of 7.3 means that this model reproduces the measured observables at the same level as the limb-darkening model reproduced for Vega.
Compared with the oblateness improved $\chi'^2$ from 1.5 to 0.95, a spot on the circular limb darkened disk improved $\chi^2$ from 150 to 7.3. 
It means that the surface brightness distribution is more essential than the oblateness for our data. 
A spot on the circular limb darkening disk model gave similar $\chi^2$ of squared visibility amplitudes at OB1 and OB2 to the limb darkened elliptical disk model.

\begin{figure*}
%\epsscale{1.10}
\plottwo{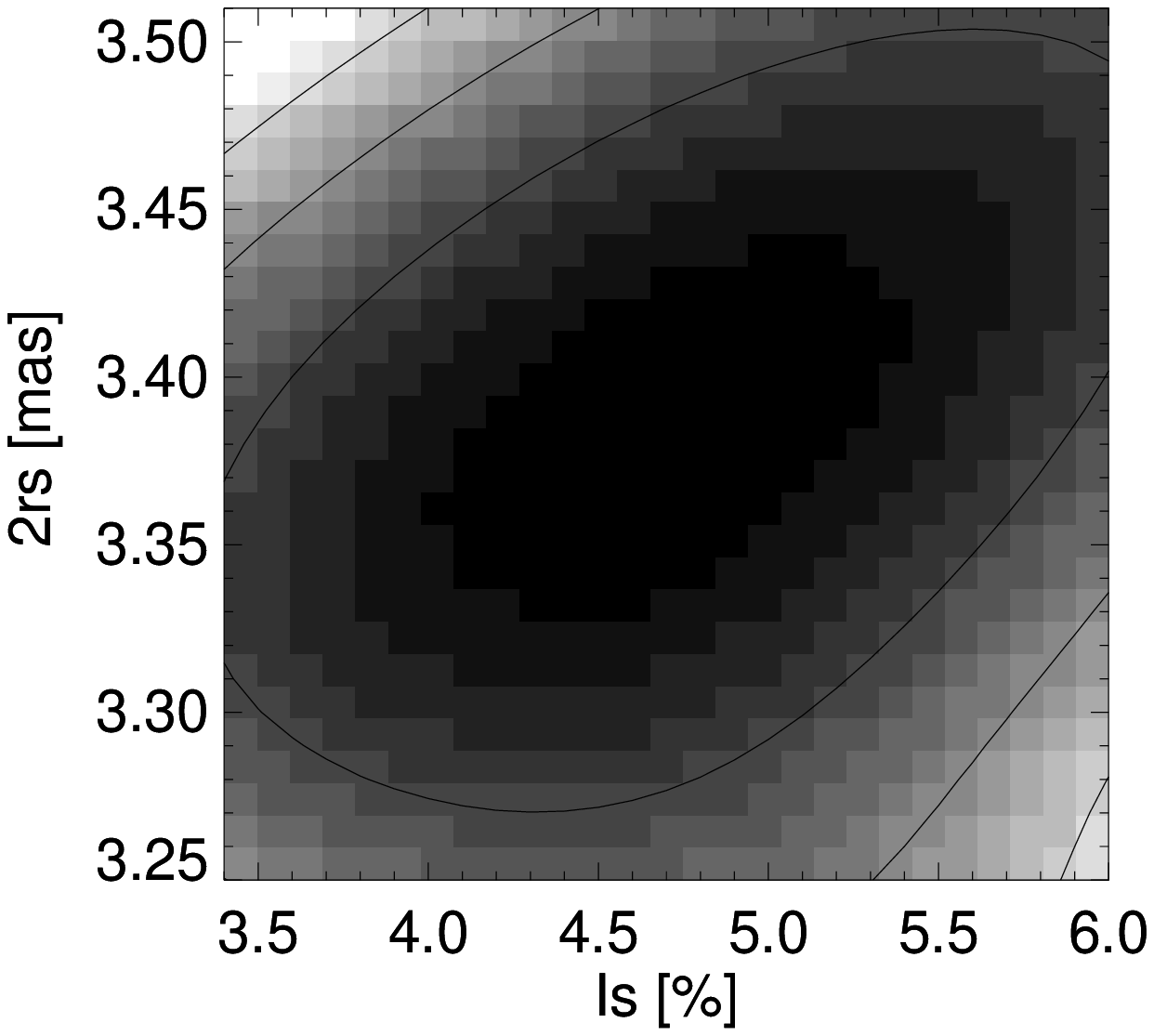}{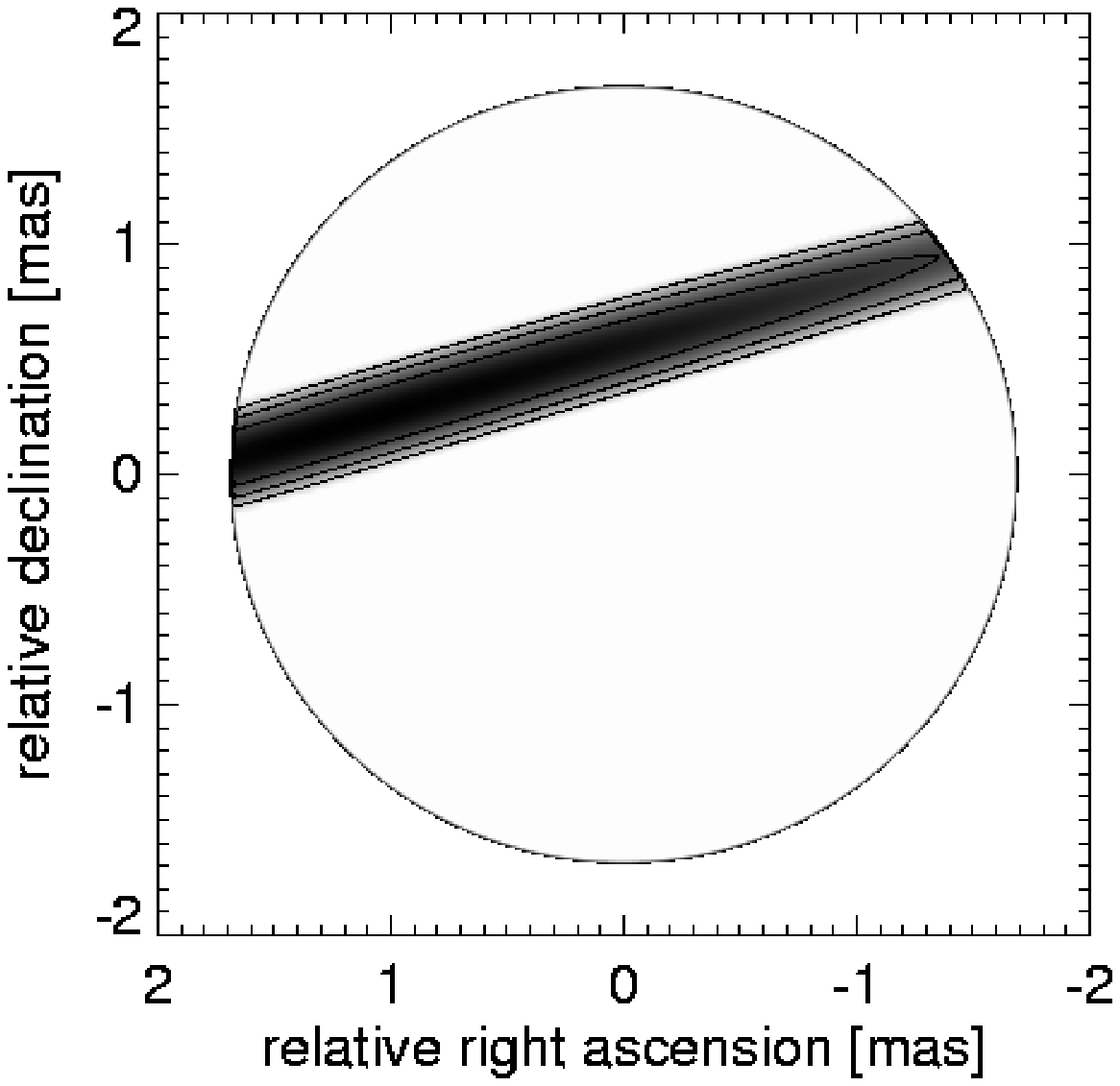}
%\vspace{-1cm}
\caption{\label{fig4} Left; $\chi^2$ map as a function of relative intensity of the bright spot, $I_{\rm p}$, and the limb darkened disk diameter, $2r_{\rm s}$. The minimum value of $\chi^2$, $\chi^2_{\rm c}$=7.3, is given for $I_{\rm p}=4.7\%$ and $2r_{\rm s}$=3.38mas. Lines show 2$\chi^2_{\rm c}$, 3$\chi^2_{\rm c}$, and 4$\chi^2_{\rm c}$. Right; $\chi^2$ map as a function of position of the bright spot, ($r_{\rm p}$, $\phi_{\rm p}$), when parameters, $I_{\rm s}$=0.047 and $2r_{\rm s}$=3.38mas, are given.}
\end{figure*}

Figure \ref{fig4} shows the $\chi^2$ map of ($I_{\rm p}$, 2$r_{\rm s}$) and the $\chi^2$ map of $(r_{\rm p}, \phi_{\rm p})$ at ($I_{\rm p}$=0.047, 2$r_{\rm s}$=3.38mas). 
Though the parameters ($I_{\rm p}$, $2r_{\rm s}$) converge, $r_{\rm p}$ and $\phi_{\rm p}$ are not determined independently. 
These two parameters appear in the second term of eq. (\ref{eq:s5_model1}).  
The equation indicates that small change of %$\phi_{\rm pa}$ 
position angle at OB3 during the observation make it difficult to determine $r_{\rm p}$ and $\phi_{\rm p}$ independently. 
Consequently, measurement of this star with wider range of position angle, or with additional long baseline which is located perpendicular to the OB3, will be needed to determine the position of the bright region accurately. 

\section{Discussion}
\label{sec5}
\begin{figure*}
%\epsscale{1.10}
\plottwo{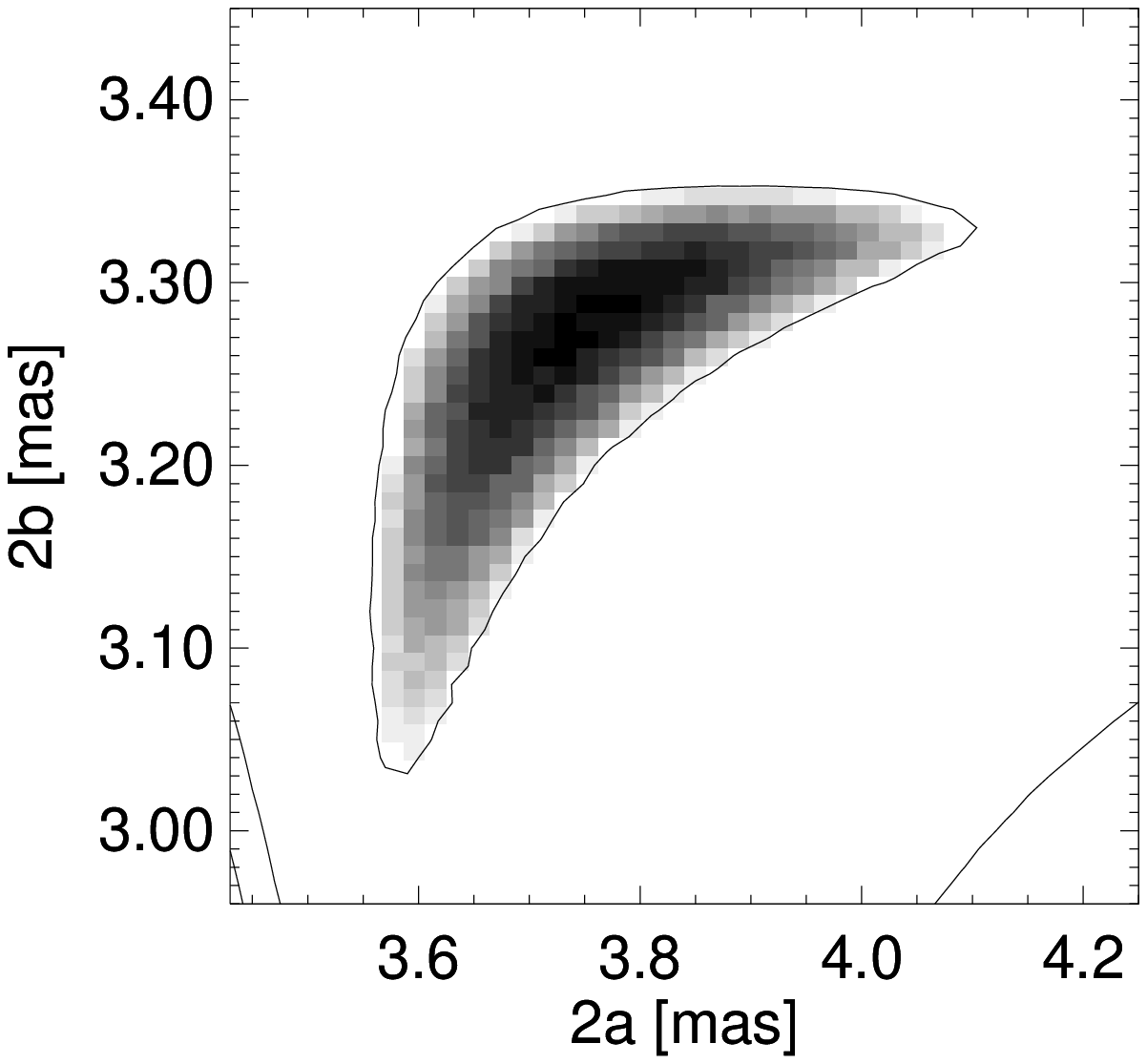}{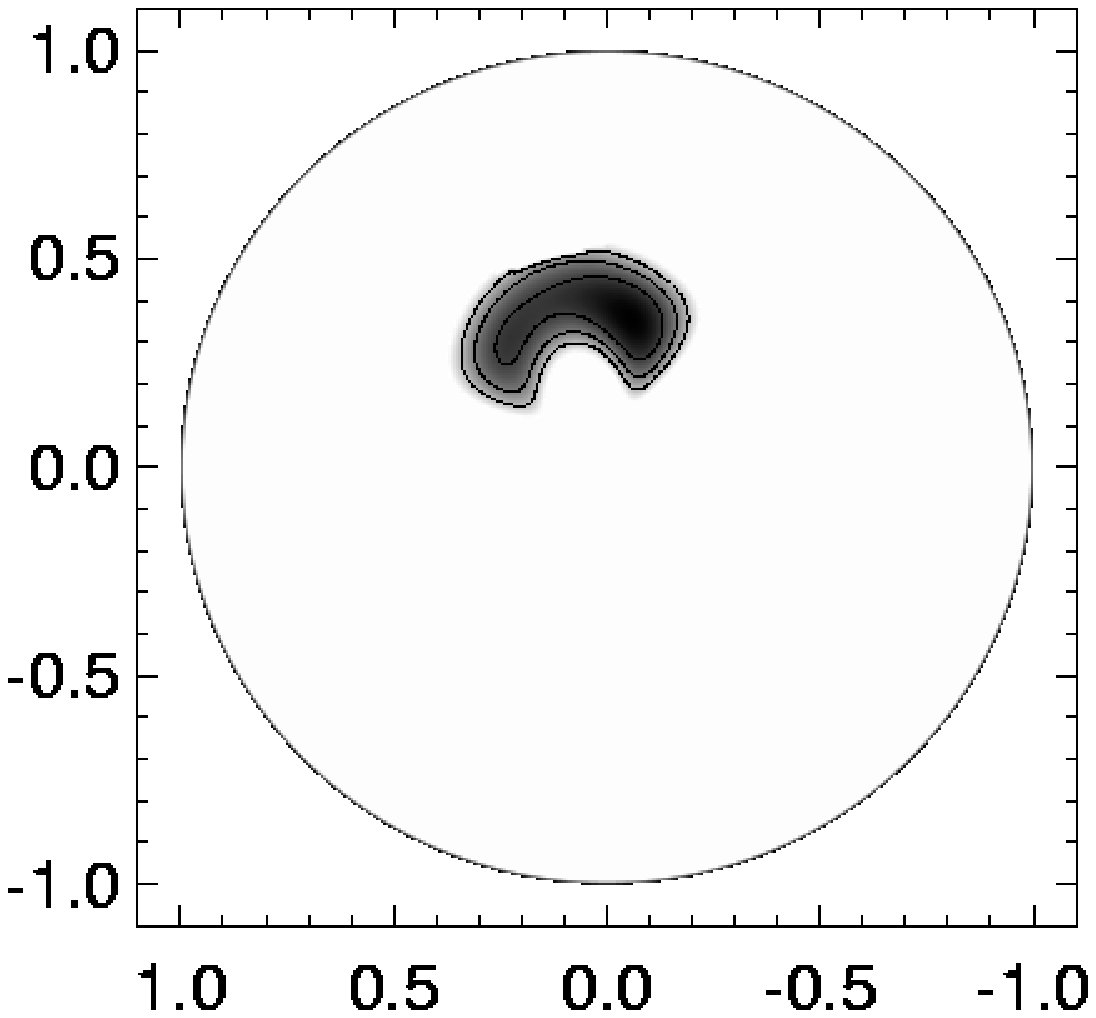}
%\vspace{-1cm}
\caption{\label{fig5} Left; $\chi^2$ map as a function of angular diameters at the major and the minor axes of ellipse, $2a$, $2b$. The minimum value of $\chi^2$, $\chi^2_{\rm e}$=7.0, is given for $2a$=3.77mas and $2b$=3.29mas. Lines show $\chi^2_{\rm c}$, 2$\chi^2_{\rm e}$, and 3$\chi^2_{\rm e}$. Right; $\chi^2$ map as a function of position of the bright spot, ($r_{\rm p}/b$, $\phi_{\rm p}$), when parameters $2a$=3.77mas and $2b$=3.29mas are given. Lines show 2$\chi^2_{\rm e}$, 3$\chi^2_{\rm e}$, and 4$\chi^2_{\rm e}$.}
\end{figure*}

The measured observables are well reproduced with a bright spot on the circular limb-darkened disk model.
Because Altair is a single star \citep{GTvB2001} and well known as a rapidly rotating star, it is natural to consider this spot is a bright pole of the gravity darkened star. 
Van Belle et al. (2001) simulated the effect of the gravity darkening as an additional 25\% brighter region covering 20\% of the surface of the star on the limb-darkened disk. % based on the calculation by Jordahl (1972). 
The relative intensity of the additional bright region is 5\% which is not far off from our result of 4.7\%. 
Considering it as a gravity-darkened star, we expect that it is better to take the effect of oblateness in addition to the surface brightness distribution into account. 
%of interest to test whether the oblateness of the stellar disk changes this result because we expect that the oblateness will add a weak constraint to the position of $\phi_{\rm p}$. 
%In former section, we tried a model of a circular limb-darkened disk, 
%
Then we replaced $r_{\rm s}$ in eq. (\ref{eq:s5_model1}) with $r_{\rm el}(\phi)$, eq. (\ref{eq:s3_rphi}), and fitted measured observables with the model, where a bright spot is on the minor axis of a limb-darkened elliptical disk. 
%
%(a little bit better model)
%If one see the rotationally symmetric elliptical stellar body with minor axis $c$ and major axis $a$ with inclination $i$, parameters of apparent shape is described by eq. (\ref{eq:s3_rphi}). 
%
Fixing the relative intensity of the spot, $I_{\rm p}$=0.047, we searched the optimized set of $(r_{\rm p}/b$, $\phi_{\rm p})$ for each set of (2$a$; 3.43$\sim$4.25mas, $\Delta 2a$=0.02mas, 2$b$; 2.96$\sim$3.45mas, $\Delta 2b$=0.01mas). 
We found the minimum $\chi^2$ decreases only a little to 7.0, which is given with the set of parameters, $2a$=3.77mas, and $2b$=3.29mas. 
Figure \ref{fig5} shows the $\chi^2$ map of ($2a$, $2b$) and the $\chi^2$ map of $(r_{\rm p}/b, \phi_{\rm p})$ at ($2a$=3.77mas, $2b$=3.29mas). 
It is worth to consider the region where $(2a, 2b)$ gives smaller $\chi^2$ than $\chi^2_{c}$ because the number of parameters are increased. 
Two parameters, ($r_{\rm p}/b$, $\phi_{\rm p}$), converge better than in the former model when parameters (2$a$, 2$b$) are given. 

Though the decrease of $\chi^2$ is small, we try to consider physical parameters of this star with the optimized set of parameters of $2a$=3.77mas, $2b$=3.29mas, $r_{\rm p}/b=0.36$, $\phi_{\rm p}=9^{\circ}$. 
Assuming a rotationally symmetric elliptical stellar body, we can calculate the inclination, $i=35^{\circ}$ using the equation, 
%
%\begin{equation}
%\label{eq:s5_c}
%c=\frac{r_{\rm p}}{\sqrt{1-\frac{b^2}{a^2}(1-\frac{r^2_{\rm p}}{b^2})}}
%\end{equation}
%
\begin{equation}
\label{eq:s5_sini}
\sin i=\sqrt{1-\frac{b^2}{a^2}(1-\frac{r^2_{\rm p}}{b^2})}. 
\end{equation}
In this case, the critical velocity of $v_{\rm c}$=430km/s \citep{DFG1976} multiplied by $\sin i$=0.58 becomes 250km/s. The values $v\sin i$=190km/s (Carpenter et al. 1984), 250km/s \citep{TRS1968} determined from spectroscopy are close to but don't break the $v_{\rm c} \sin i$. 
The apparent brightness distribution with these parameters are shown in Fig. \ref{fig7}. 
\begin{figure*}
%\epsscale{.80}
\plotone{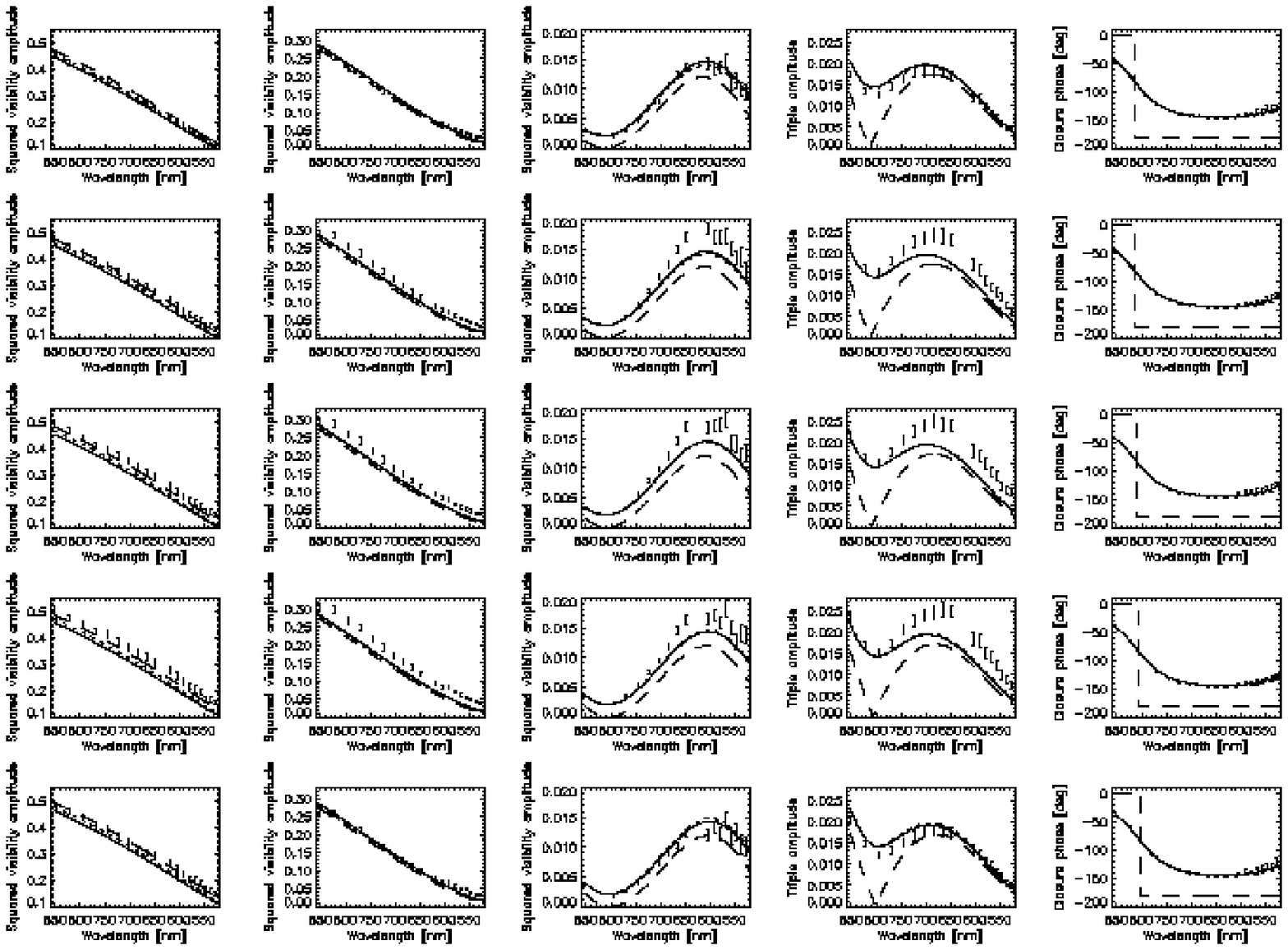}
\plotone{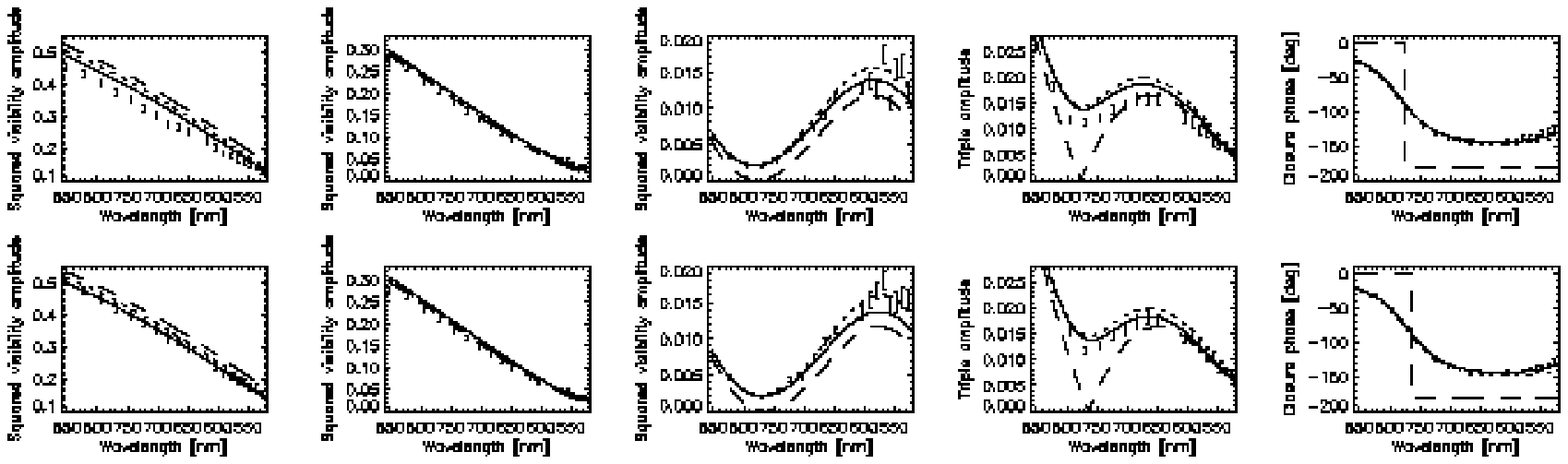}
\vspace{-6cm}
\caption{\label{fig6} Measured observables, squared visibility amplitudes, triple amplitudes and close phases of Altair with models. Solid lines show a spot on the elliptical limb-darkened disk model. Short dashed lines show a spot on the circular limb-darkened disk model. Dashed lines show limb-darkened disk model without spot. }
\end{figure*}
\begin{figure}
\epsscale{.80}
\plotone{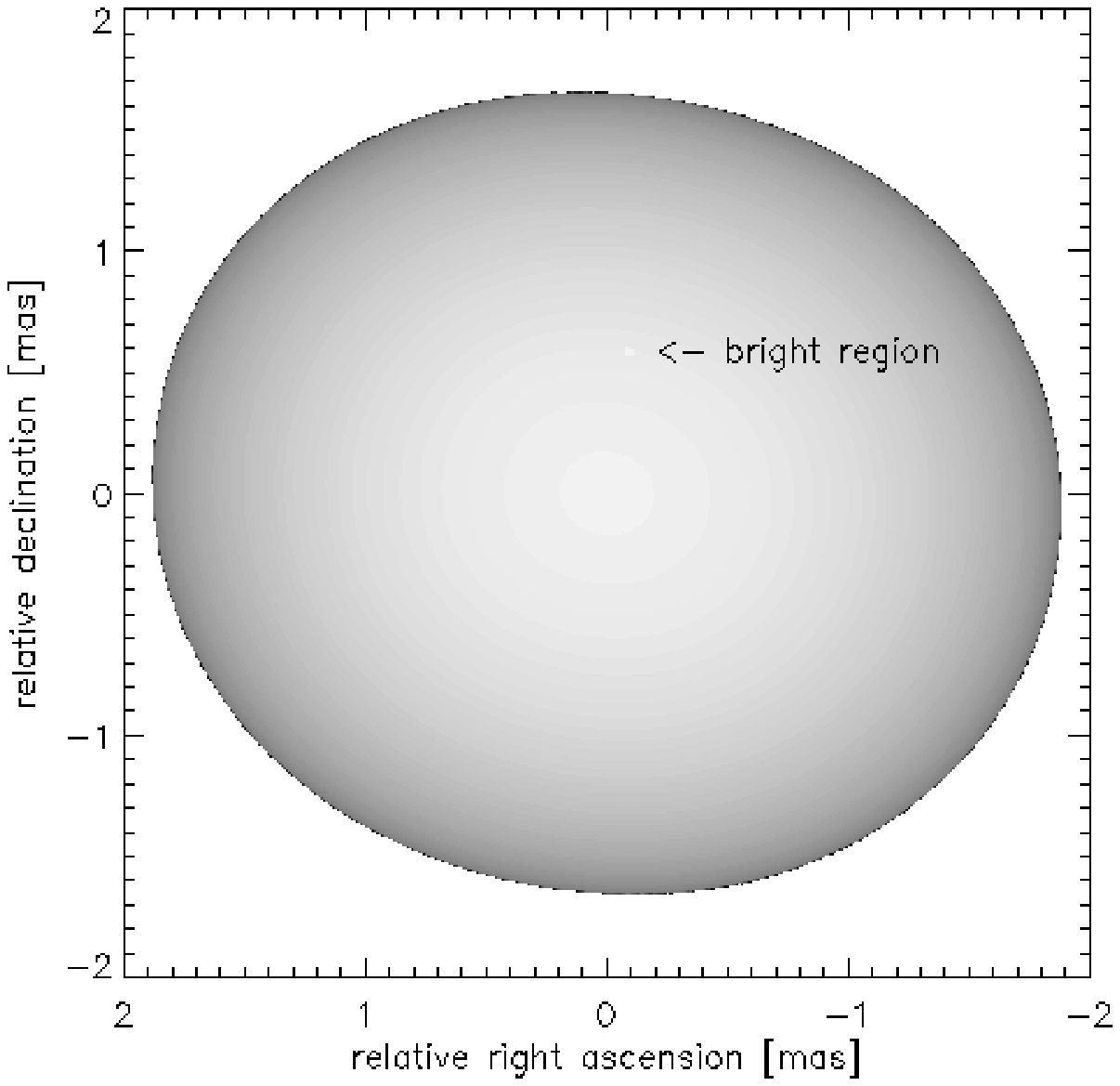}
\caption{\label{fig7} A possible solution of a bright spot on the elliptical limb darkening disk model of Altair projected onto the sky with parameters; angular diameter at the major axis $2a=3.77$mas, angular diameter at the minor axis $2b=3.29$mas, orientation of the rotational axis projected onto the sky from the north pole $\phi_0=9^{\circ}$, realtive intensity of the bright spot $I_{\rm p}=4.7\%$, inclination $i=35^{\circ}$. }
\end{figure}

\section{Conclusion}
\label{sec6}
We observed Altair with high resolution including the measurement of the triple product using three long baselines of the NPOI for four nights. 
Measured observables indicate the asymmetric surface brightness distribution of this star; the asymmetry is deduced directly from the definition of the visibility and does not depend on model. 
Though the measurement of the structure of the surface brightness distribution has been reported for evolved stars \citep{PGT1997}, this is the first time that the asymmetric surface brightness distribution for a main sequence star has been found from the direct measurement using interferometry. 
The measured observables are well reproduced with a model of a bright spot of relative intensity 4.7\% on the limb-darkening disk with an angular diameter of 3.38mas. 
The rapid rotation of the star indicates that the bright spot is a bright pole of gravity-darkened star. 
Though we couldn't determine the position of the pole owing to insufficient data, we expect that additional observation with sets of long baselines covering wider range of position angle will solve this problem. 
More sophisticated gravity and limb darkening modeling of a rapidly rotating star \citep{ADdS2002} will also help the determination of physical parameters when we have sufficient multi baselines data. 
We expect these kinds of observations will be realized with current and future interferometers and that study on rapidly rotating stars will progress with further observation by interferometers.

\acknowledgments

N. Ohishi acknowledges C. A. Hummel for supporting the use of data reduction system, OYSTER. 
The Navy Prototype Optical Interferometer is a joint project of the Naval Research Lab and the US Naval Observatory in cooperation with Lowell Observatory, and is funded by the Office of Naval Research and the Oceanographer of the Navy.


\begin{thebibliography}{}
\bibitem[Armstrong et al. 1998]{AJT1998} 
Armstrong, J. T., Mozurkewich, D., Rickard, L. J., 
Hutter, D. J., Benson, J. A., Bowers, P. F., Elias II, N. M., 
Hummel, C. A., Johnston, K. J., Buscher, D. F., Clark III, J. H., 
Ha, L., Ling, L. -C., White, N. M., and Simon, R. S. 
1998, \apj, 496, 550 %-571
%
\bibitem[Benson et al. 1997]{JAB1997} 
Benson, J. A., Hutter, D. J., Elias II, N. M., Bowers, P. F., Johnston, K. J., 
Hajian, A. R., Armstrong, J. T., Mozurkewich, D., Pauls, T. A.,
Rickard, L. J., Hummel, C. A., White, N. M., Black, D., and Denison, C. S. 
1997, \aj, 114, 1221 %-1226
%
\bibitem[Carpenter et al. 1984]{KGC1984}
Carpenter, K. G., Slettebak, A., and Sonneborn, G. 
1984, \apj, 286, 741 %-746
%rotational velocities of later B type and A type stars as determined from
%ultraviolet versus visual line profiles
%
\bibitem[Ciardi et al. 2001]{DRC2001}
Ciardi, D. R., van Belle, G. T., Akeson, R. L., Thompson, R. R., 
Lada, E. A., and Howell, S. B. 
2001, \apj, 559, 1147 %-1154 
%
\bibitem[Claret 2000]{AC2000}
Claret, A. 2000, \aap, 359, 289 %-298
%Studies on stellar rotation 
%II. Gravity-darkening: the effects of the input physics 
%and differential toration. New results for very low mass stars 
%
\bibitem[Colavita et al. 1999]{MMC1999} 
Colavita, M. M., Wallace, J. K., Hines, B. E., Gursel, Y., 
Malbet, F., Palmer, D. L., Pan, X. P., Shao, M., Yu, J. W., 
Boden, A. F., Dumont. P. J., Gubler, J., Koresko, K., Kulkarni, S. R., 
Lane, B. F., Mobley, D. W., and van Belle, G. T. 
1999, \apj, 510, 505 %-521
%
\bibitem[Domiciano de Souza et al. 2002]{ADdS2002} 
Domiciano de Souza, A., Vakili, F., Jankov, S., Janot-Pacheco, E., and Abe, L. 
2002, \aap, 393, 345 %-357
%
\bibitem[Domiciano de Souza et al. 2003]{ADdS2003} 
Domiciano de Souza, A., Kervella, P., Jankov, S., 
Abe, L., Vakili, F., di Folco, E., and Paresce, F. 
2003, \aap, 407, L47 %-L50
%The spinning-top Be star Achernar from VLTI-VINICI
%
\bibitem[Freire Ferrero et al. 1983]{RFF1983}
Freire Ferrero, R., Gouttebroze, P., Catalano, S., Marilli, E., 
Bruhweiler, F., Kondo, Y., van der Hucht, K., and Talavera, A. 
1983, \apj, 121, 59 %-68
%
\bibitem[Gray 1976]{DFG1976}
Gray, D. F., 1976, The Observation and Analysis of Stellar Photospheres 
(New York; Wiley-Interscience)
%
%\bibitem[Lucy 1967]{LBL1967}
%Lucy, L. B. 1967, ZA, 65, 89 %-92
%
\bibitem[Hanbury Brown et al. 1967]{RHB1967}
Hanbury Brown, R., Davis, J., Allen, L. R., and Rome J. M. 
1967, \mnras, 137, 393 %-417
%The stellar interferometer at narrabri observatory II
%
\bibitem[Hummel et al. 1998]{CAH1998} 
Hummel, C. A., Mozurkewich, D., Armstrong, J. T., 
Hajian, A. R., Elias II, N. M., and Hutter, D. J. 
1998, \aj, 116, 2536 %-2548
%
%\bibitem[Hajian et al. 1998]{ARH1998} 
%Hajian, A. R., Armstrong, J. T., Hummel, C. A., Benson, J. A., 
%Mozurkewich, D., Pauls, T. A., Hutter, D. J., Elias II, N. M., 
%Johnston, K. J., Rickard, L. J. and White, N. M. 
%1998, \apj, 496, 484 %-489 
%
%\bibitem[Jordahl, 1972]{PRJ1972}
%Jordahl, P. R. 1972, Ph. D. thesis, Univ. Texas, Austin 
\bibitem[Mozurkewich et al. 1991]{DM1991}
Mozurkewich, D., Johnston, K. J., Simon, R. S., Bowers, P. F., and Gaume, R. 
1991, \aj, 101, 2207 %-2219 
%
%\bibitem[Quirrenbach et al. 1994]{AQ1994}
%Quirrenbach, A., Buscher, D. F., Mozurkewich, D., 
%Hummel, C. A., and Armstrong, J. T. 
%1994, \aap,  283, L13 %-L16
%Maximum-entropy maps of the Be shell star zeta Tauri 
%from optical long-baseline interferometry 
%
\bibitem[Nordgren et al. 1999]{TEN1999}
Nordgren, T. E., Germain, M. E., Benson, J. A., Mozurkewich, D., Sudol, J. J., Elias II, N. M., Hajian, A. R., White, N. M., Hutter, D. J., Johnston, K. J., Gauss, F. S., Armstrong, J. T., Pauls, T. A., and Rickard, L. J. 
1999, \aj, 118, 3032 %-3038
%
%
\bibitem[Nordgren,Sudol,and Mozurkewich 2001]{TEN2001}
Nordgren, T. E., Sudol, J. J., Mozurkewich, D. 
2001, \aj,  122, 2707 %-
%Maximum-entropy maps of the Be shell star zeta Tauri 
%from optical long-baseline interferometry 
%
\bibitem[Quirrenbach et al. 1996]{AQ1996} 
Quirrenbach, A., Mozurkewich, D., Buscher, D. F., 
Hummel, C. A., and Armstrong, J. T. 
1996, \aap, 312, 160 %-166
%
\bibitem[Royer et al. 2002]{FR2002}
Royer, F., Grenier, S., Baylac, M.-O., Gomez, A. E., and Zorec, J. 
2002, \aap, 393, 897 %-911
%
\bibitem[Stoeckley 1968]{TRS1968}
%T. R. Stoeckley, 
Stoeckley, T. R. 
1968, \mnras, 140, 121 %-139
%
\bibitem[Tuthill,Haniff,and Baldwin 1997]{PGT1997}
%\bibitem[Tuthill,Haniff 1997]{PGT1997}
%P. G. Tuthill, C. A. Haniff, and J. E. Baldwin, 
Tuthill, P. G., Haniff, C. A., and Baldwin, J. E. 
1997, \mnras, 285, 529 %-539
%Hotspots on late-type supergiants
%
\bibitem[von Zeipel 1924]{vZ1924}
%H. von Zeipel, 1924, \mnras, 84, 665 %
von Zeipel, H. 1924, \mnras, 84, 665 %
%
\bibitem[van Belle et al. 2001]{GTvB2001} 
%G. T. van Belle, D. R. Ciardi, R. R. Thompson, R. L. Akeson and E. A. Lada, 
van Belle, G. T., Ciardi, D. R., Thompson, R. R., Akeson, R. L., 
and Lada, E. A. 
2001, \apj, 559, 1155 %-1164
%
\bibitem[Van Hamme 1993]{WVH1993} 
%W. Van Hamme, 
Van Hamme, W. 
1993, \aj, 160, 2096 %-2117
%
\bibitem[Wittkowski et al. 2001]{MW2001} 
Wittkowski, M., Hummel, C. A., Johnston, K. J., Mozurkewich, D., 
Hajian, A. R., and White, N. M. 
2001, \aap, 377, 981 %-993
%
\end{thebibliography}
\end{document}